\documentclass[prd,twocolumn,tightenlines,showpacs,nofootinbib]{revtex4-1} 

\usepackage{amsfonts}
\usepackage{dcolumn}
\usepackage{bm}
\usepackage{amsmath}
\usepackage{nicefrac}
\usepackage{amsthm}
\usepackage{amssymb}
\usepackage{graphicx}
\usepackage{color}

\newcommand{\appropto}{\mathrel{\vcenter{
  \offinterlineskip\halign{\hfil$##$\cr
    \propto\cr\noalign{\kern2pt}\sim\cr\noalign{\kern-2pt}}}}}

\begin{document}

\title{Searches for topological defect dark matter via non-gravitational signatures}

\date{\today}
\author{Y.~V.~Stadnik} \email[]{y.stadnik@unsw.edu.au}
\affiliation{School of Physics, University of New South Wales, Sydney 2052,
Australia}
\author{V.~V.~Flambaum}
\affiliation{School of Physics, University of New South Wales, Sydney 2052,
Australia}

\begin{abstract}

We propose schemes for the detection of topological defect dark matter using pulsars and other luminous extraterrestrial systems via non-gravitational signatures. The dark matter field, which makes up a defect, may interact with standard model particles, including quarks and the photon, resulting in the alteration of their masses. When a topological defect passes through a pulsar, its mass, radius and internal structure may be altered, resulting in a pulsar `quake'. A topological defect may also function as a cosmic dielectric material with a distinctive frequency-dependent index of refraction, which would give rise to the time delay of a periodic extraterrestrial light or radio signal, and the dispersion of a light or radio source in a manner distinct to a gravitational lens. A topological defect passing through Earth may give rise to temporary non-zero electric dipole moments for an electron, proton, neutron, nuclei and atoms.


\end{abstract}




\maketitle 

\emph{Introduction.} --- 
Dark matter remains one of the most important unsolved problems in physics \cite{Beringer2012PDG}. Apart from the usual uniformly distributed dark matter, the existence of stable, extended-in-space configurations of dark matter, which may have formed at early cosmological times, are also possible. Such dark matter configurations are generally termed as topological defect dark matter (to which we shall refer to in this work as simply `defects') and can have various dimensionalities: 0D (corresponding to monopoles), 1D (strings) and 2D (domain walls) \cite{Vilenkin1985}. The transverse size $d$ of a defect cannot be predicted from existing theory in an \emph{ab initio} manner, but typically scales as $d \propto 1/m_\phi$, where $m_\phi$ is the mass of the particles making up the defect. Defects have primarily been sought for via their gravitational effects, including gravitational lensing and gravitational radiation (see e.g.~Refs.~\cite{TDDM_Book1,TDDM_Book2,TDDM_Book3} and the plethora of references within). Constraints on the contribution of cosmic defects to observed temperature fluctuations in the CMB spectrum have been placed by recent results from Planck \cite{Planck2013-Defects} and BICEP2 \cite{BICEP2014,Moss2014-BICEP2}, but the existence of cosmic defects is neither confirmed nor not ruled out by these results, leaving the tantalising question of whether or not defects exist still unanswered.

In more recent times, several schemes have been proposed for the detection of non-gravitational effects induced by a defect passing directly through Earth. Ref.~\cite{Pospelov2013} proposes that a global network of magnetometers is used to directly search for the interaction of the form $\mathbf{\sigma} \cdot \mathbf{B}_{\textrm{eff}}$, induced by an axion-like pseudoscalar domain wall temporarily passing through Earth. Ref.~\cite{Derevianko2013} proposes to use a global network of synchronised atomic clocks for defect detection. The passage of a defect, inside which there may be variation of fundamental constants induced by the interaction of the constituent dark matter, which makes up the defect, with standard model (SM) particles, through this network of clocks would alter the transition frequencies in (and hence the time recorded by) different clocks as the defect passes into, then out of Earth. 


In the present work, we propose schemes for the detection of topological defect dark matter using pulsars and other luminous extraterrestrial systems via non-gravitational signatures. The biggest advantage of such astrophysical observations over the proposed terrestrial detection methods of Refs.~\cite{Pospelov2013,Derevianko2013} is the much higher probability of a defect been found in the vast volumes of outer space compared with one passing through Earth itself. Pulsars are highly magnetised, rotating neutron stars with periods ranging from $T = 1.5$ ms $-$ $8.5$ s \cite{Stairs2003Alive}. The stabilities of pulsars as timekeeping devices are of the order $\sim 10^{-15}$ \cite{Lyne2012Pulsar}, which are second only to terrestrial atomic clocks, the best stabilities of which are presently of the order $\sim 10^{-18}$ \cite{Hinkley2013,Bloom2014}. The dark matter field, which makes up a defect, may interact with SM particles, including quarks and the photon, resulting in the alteration of their masses. When a topological defect passes through a pulsar, its mass and equilibrium radius may be altered, due to alterations in the mass of a neutron inside a defect. This process may also be accompanied by a significant change in the pulsar's internal structure (including the unpinning of quantised vortices, which carry angular momentum inside a pulsar's core), which is believed to consist of at least one superfluid component, and internal dynamics, leading to a non-equilibrium state, which undergoes slow relaxation and may be one of the possible explanations for the pulsar glitch phenomenon (see e.g.~Refs.~\cite{Baym1969,Anderson1975,Alpar1984,Pines1985,Takatsuka1988,Ravenhall1994,Link1999,Andersson2003,Chamel2005,Donati2006,Melatos2007,Glampedakis2009,Van2010pulsar,Pizzochero2011,Andersson2012pulsar,Chamel2012,Steiner2014glitchcrust,Piekarewicz2014,Li2014glitches}).
At present, pulsar glitches are not well understood and theories based on internally-driven mechanisms of pulsar glitches raise a number of questions, which challenge existing understanding of the pulsar glitch phenomenon, see e.g.~Ref.~\cite{Andersson2012pulsar}. 
The net result of defect passage through a pulsar is thus a pulsar `quake' and an altered frequency of rotation of a pulsar. From existing pulsar glitch data \cite{Pulsar_CSIRO_Database}, in which the frequency of rotation of a pulsar is observed to increase abruptly, typically in the range $\delta \omega / \omega = 10^{-11} - 10^{-5}$, with a comparatively slow recovery period, which generally ranges from $T_d = 1$ day $-$ $3$ years, we infer that the upper limit range of neutron mass variations inside defects is $\delta m_n / m_n \sim 10^{-11} - 10^{-5}$. We discuss the implications of this inference for the detection of defects using a global network of atomic clocks, which is proposed in \cite{Derevianko2013}. 

A topological defect may also function as a cosmic dielectric material with a frequency-dependent index of refraction, which would give rise to the time delay of a periodic extraterrestrial light or radio signal, such as that from a background pulsar, and the dispersion of an extraterrestrial light or radio source in a manner distinct to a gravitational lens. Such lensing is distinct from the conventionally sought gravitational lensing of background radiation by a defect, which is a frequency-independent effect. A defect, which consists of axion-like pseudoscalar dark matter, passing through Earth may give rise to temporary non-zero electric dipole moments for an electron, proton, neutron, nuclei and atoms.

\emph{Theory.} --- 
A scalar dark matter field $\phi$ may interact with fermions via the coupling \cite{Derevianko2013}:
\begin{equation}
\label{L_Varn_FCs}
\mathcal{L}_{\textrm{int}}^{f} = - \sum_{f=e,p,n,...} m_f \left(\frac{\phi c}{\Lambda_f}\right)^2 \bar{\psi}_f \psi_f ,
\end{equation}
where $\psi_f$ is the fermion Dirac field, $\bar{\psi}_f = \psi_f^{\dagger} \gamma^0$ is the corresponding fermion adjoint field, $m_f$ is the standard mass of the fermion, $\Lambda_f$ is the reciprocal of the coupling constant for the interaction with a particular fermion, and the sum is over the fermions $f$. The coupling (\ref{L_Varn_FCs}) leads to the following alterations in the fermion masses:
\begin{equation}
\label{delta_mfs}
m_f \to m_f \left[1 + \left(\frac{\phi}{\Lambda_f} \right)^2 \right] .
\end{equation}
The quadratic dependence on $\phi$ in (\ref{L_Varn_FCs}) in the form $\phi^2/\Lambda_f^2$, rather than linear dependence on $\phi$ in the form $\phi/\Lambda_f$, allows one to escape from the very strong constraints imposed by the nil results of searches for the fifth force and the violation of the equivalence principle \cite{Olive2008P}. Both direct laboratory and astrophysical constraints on $\Lambda_f$ do not exceed $\sim 10$ TeV \cite{Derevianko2013}.

A scalar dark matter field might also interact with a photon via the following coupling:
\begin{equation}
\label{L_Varn_FCs2}
\mathcal{L}_{\textrm{int}}^{\gamma} = \frac{1}{8\pi} \left(\frac{\phi c}{\Lambda_\gamma \hbar}\right)^2 A^\nu A_\nu ,
\end{equation}
where $A^{\nu}$ is the photon field and $\Lambda_\gamma$ is the reciprocal of the coupling constant for the given interaction. The coupling (\ref{L_Varn_FCs2}) necessitates the choice of gauge: $\partial_\nu \left( \phi^2 A^\nu \right) = 0$ in flat spacetime, which follows from the application of Lagrange's field equations, and is similar to the coupling in Proca theory:
\begin{equation}
\label{L_Proca-mass}
\mathcal{L}_{\textrm{Proca}} = \frac{1}{8\pi} \left(\frac{m c}{\hbar}\right)^2 A^\nu A_\nu ,
\end{equation}
which gives rise to a massive vector boson with mass $m$. Comparing (\ref{L_Varn_FCs2}) and (\ref{L_Proca-mass}) shows that the interaction (\ref{L_Varn_FCs2}) results in a photon with mass $m_{\gamma} = \phi/\Lambda_\gamma$. Assuming that the speed of a defect is $v_{\textrm{TD}} \ll c$, the energy of a photon may thus be taken to be unaltered upon its passage from vacuum into a defect, implying that the speed of a photon inside a defect is given by:
\begin{equation}
\label{v_photon_TD}
v_{\gamma} = c \sqrt{1 - \left( \frac{\phi c^2}{\hbar \omega \Lambda_\gamma} \right)^2} ,
\end{equation}
with the index of refraction inside the defect being:
\begin{equation}
\label{n_IR_TD}
n (\omega) = \frac{1}{\sqrt{1 - \left( \frac{\phi c^2}{\hbar \omega \Lambda_\gamma} \right)^2}} .
\end{equation}
Thus a defect may function as a cosmic dielectric material with a distinctive frequency-dependent index of refraction.

We note that interaction (\ref{L_Varn_FCs2}) is the simplest mechanism, through which the photon mass is altered. There may be more complicated gauge-invariant Lagrangians, which alter the photon dispersion relation.

\emph{Pulsar quakes.} --- 
Consider a defect passing directly through a pulsar itself. There is friction acting between the defect and pulsar, which reduces the angular momentum of the pulsar $L = I \omega$, where the moment of inertia of the pulsar is $I \sim \frac{2}{5} M R^2$, albeit most likely only very slightly and so we ignore the effects of friction in the ensuing discussion. Thus we can write small relative changes in the frequency of rotation of a pulsar as:
\begin{equation}
\label{del-omega_on_omega}
\frac{\delta \omega (t)}{\omega} = - \frac{\delta M (t)}{M} - \frac{2 \delta R (t)}{R} .
\end{equation}
The interaction (\ref{L_Varn_FCs}), which alters the fermion masses according to (\ref{delta_mfs}), effectively increases the mass of a pulsar when the pulsar is immersed in a defect. A less obvious implication of (\ref{delta_mfs}) is that the equilibrium radius of a pulsar is decreased when the pulsar is immersed in a defect. This can be seen from the requirement of hydrostatic equilibrium, in which the outward pressure exactly balances the inward gravitational pressure. The abrupt change in a pulsar's mass and equilibrium radius may also be accompanied by a significant change in the pulsar's internal structure and dynamics (including the unpinning of quantised vortices, which carry angular momentum inside a pulsar's core), leading to a non-equilibrium state, which undergoes slow relaxation and may be one of the possible explanations for the pulsar glitch phenomenon \cite{Baym1969,Anderson1975,Alpar1984,Pines1985,Takatsuka1988,Ravenhall1994,Link1999,Andersson2003,Chamel2005,Donati2006,Melatos2007,Glampedakis2009,Van2010pulsar,Pizzochero2011,Andersson2012pulsar,Chamel2012,Steiner2014glitchcrust,Piekarewicz2014,Li2014glitches}. 

The neutron equation-of-state in extremely dense enviroments, such as those found inside a pulsar, is not known precisely (see e.g.~Refs.~\cite{Brown2000neutron,Typel2001neutron,Steiner2005isospin,Steiner2013neutron,Brown2013constraints,Brown2014constraints}), so we make use of the simplest possible model: a non-relativistic degenerate neutron gas. Due to the simplicity of our model, the result obtained from the ensuing analysis should be considered an estimate only. The pressure of a degenerate neutron gas is given by \cite{LL5}:
\begin{equation}
\label{P_deg-Fermi-gas}
P_d = \frac{(3 \pi^2)^{2/3}}{5} \frac{\hbar^2}{m_n} \left(\frac{N}{V}\right)^{5/3} ,
\end{equation}
where $N$ is the number of neutrons and $V$ is the volume of the system. The gravitational self-energy of a uniform sphere of mass $M$ and radius $R$ is:
\begin{equation}
\label{U_grav_self-energy}
U_g = - \frac{3GM^2}{5R} ,
\end{equation}
and so the inward gravitational pressure is:
\begin{equation}
\label{P_grav}
P_g = -\frac{\partial U_g}{\partial V} = - \frac{3GM^2}{20\pi R^4} .
\end{equation}
Equating the pressures (\ref{P_deg-Fermi-gas}) and (\ref{P_grav}) gives the radius of a pulsar:
\begin{equation}
\label{R_pulsar_star}
R = \frac{3 \left(\frac{3}{2} \right)^{1/3} \hbar^2 \pi^{2/3}}{2 GM^{1/3} m_n^{8/3}} ,
\end{equation}
from which we arrive at the following change in the radius of a pulsar:
\begin{align}
\label{delta-R0_eqm}
\frac{\delta R}{R} &= -\frac{1}{3} \frac{\delta M}{M} - \frac{8}{3} \frac{\delta m_n}{m_n} \notag \\
&\approx - 3 \frac{\delta m_n}{m_n}  , 
\end{align}
where in the second line of (\ref{delta-R0_eqm}), we have used the fact that neutrons are the dominant form of matter in a pulsar. From Eqs.~(\ref{del-omega_on_omega}) and (\ref{delta-R0_eqm}), we see that:
\begin{equation}
\label{result_p}
\delta \omega / \omega \sim \delta m_n / m_n .
\end{equation}
From observed rotational frequency variations associated with pulsar glitches \cite{Pulsar_CSIRO_Database}, we infer that the upper limit range of neutron mass variations inside defects is:
\begin{equation}
\label{result}
\delta m_n / m_n \sim 10^{-11} - 10^{-5} .
\end{equation}
We note that defect passage through a pulsar may also be accompanied by the unpinning of quantised vortices, which carry angular momentum inside a pulsar's core, resulting in an enhancement of the observed increase in pulsar rotational frequency. In addition, there may be significant differences in the internal structures and dynamics of pulsars, as well as their responses to perturbations. Thus a defect need not necessarily be larger than a pulsar in size and a single type of defect with a fixed size, in which neutron mass variations are of the order $\delta m_n / m_n \sim 10^{-11}$, for instance, may in principle explain glitches of all sizes in the range $10^{-11} - 10^{-5}$.

A more precise constraint on neutron mass variations inside a defect from existing pulsar glitch data may be obtained in principle from advanced nuclear structure and dynamics calculations, in the presence of defects. Such calculations might also yield important information regarding the types and sizes of defects that most likely explain observed pulsar glitches.



\emph{Further astrophysical observations.} --- 
We suggest further astrophysical observations for the detection of defects via non-gravitational signatures. Suppose that a defect passes through the line-of-sight connecting a pulsar and Earth. The speed of light inside the defect is given by (\ref{v_photon_TD}), with $v_\gamma < c$. The passage of a defect into and out of the line-of-sight would, therefore, result in small time delay and time advancement of pulsar signals reaching Earth, respectively. The direct observable of interest is the phase shift $\eta (t)$ in the function $\cos[\omega t + \eta (t)]$, where $\omega$ is the reference frequency. The phase shift $\eta(t_A) = 0$ prior to the passage of a defect through the line-of-sight connecting a pulsar and Earth. For light passing through a defect of length $l$ along the direction of the line-of-sight, the phase shift is: 
\begin{equation}
\label{Phi_Los}
\eta(t_B) = l \omega \left( \frac{1}{c} - \frac{1}{v_{\gamma}} \right) ,
\end{equation}
which is negative on account of time delay of the signal. Finally, $\eta(t_C) = 0$ after the defect has fully passed the line-of-sight. 

To measure the discussed phase shifts, one could use, for instance, a terrestrial atomic clock or a second pulsar to provide a reference frequency. One would then measure the phase shift of an initially synchronised pulsar/clock or pulsar/pulsar system. For examples of the wide range of clock systems, which may be used, we refer the reader to Refs.~\cite{Allan1997,Peik2004,Takamoto2005opticalSrLAttice,Oskay2006,Tobar2006Sapphire,Ludlow2008,Chou2010,Jiang2011Yb_opt_lattice,Derevianko2011C,Hinkley2013,Bloom2014}.

Now suppose that a defect passes more generally between some luminous extraterrestrial object, such as a pulsar, quasar, galaxy or star, and Earth. According to (\ref{n_IR_TD}), a defect functions as a cosmic dielectric object with a distinctive variable index of refraction. Thus a defect can lens electromagnetic radiation in a manner distinct from the gravitational lensing of electromagnetic radiation by a massive body \cite{Misner_Gravitation}. Lensing as a result of (\ref{n_IR_TD}) is due to light being scattered off a defect and thus is short-ranged in nature, whereas gravitational lensing occurs due to the curvature of space-time by a massive body, is long-ranged in nature and is primarily due to radiation passing around the gravitating body. Lensing of electromagnetic radiation by a defect functioning as a cosmic dielectric with the refractive index (\ref{n_IR_TD}) exhibits a strong dependence on the incident photon frequency, giving rise to dispersion, which is responsible for such common everyday phenomena as the Rainbow \cite{Zangwill_MED}. On the other hand, gravitational lensing of radiation by a defect is independent of the incident photon frequency, since the timelike geodesic trajectories, on which photons propagate, are determined by the curvature of space-time and hence bear no direct relation to the photon frequency. Lensing of radiation by a defect can also be sought for in conjunction with the onset of a pulsar glitch, since when a defect passes through a pulsar, the radiation emitted by a pulsar should also pass through the defect.

\emph{Parameter spaces for defect signatures.} --- 
Defect-induced variation of neutron mass may be expressed as follows with regard to defect parameters \cite{Derevianko2013}:
\begin{equation}
\label{Def_param2}
\frac{\delta m_n}{ m_n} \sim \frac{A^2}{\Lambda_n^2} = \frac{\rho_{\textrm{TDM}}v_{\textrm{TD}} \tau d}{\Lambda_n^2} ,
\end{equation}
where $A$ is the maximum amplitude of the scalar field inside a defect, $\rho_{\textrm{TDM}}$ is the energy density associated with a topological defect network, $v_{\textrm{TD}}$ is the typical speed of a defect and $\tau$ is the average time between encounters of a system with defect objects. Note that Eq.~(\ref{Def_param2}) applies to defects of all dimensionalities $n=0,1,2$. 
Assuming $\rho_{\textrm{TDM}} = \rho_{\textrm{CDM}} \approx 7.6 \times 10^{-4}$ eV$^4$, $v_{\textrm{TD}} \sim 10^{-3} c$ and $\tau \sim 1$ year (since time intervals between glitch events in pulsars generally vary from $\sim 1 - 10$ years), we find, for defect-induced neutron mass variations in the range $\delta m_n / m_n \sim 10^{-11} - 10^{-5}$, that the combination of parameters $d/\Lambda_n^2$ is in the range:
\begin{equation}
\label{fun!}
\frac{d}{\Lambda_n^2} \sim (10^{-27} - 10^{-21}) ~ \textrm{eV}^{-3} .
\end{equation}
Combined with the laboratory and astrophysical constraints $\Lambda_n \gtrsim 10$ TeV \cite{Derevianko2013}, (\ref{fun!}) points to defects with transverse sizes in the range:
\begin{equation}
\label{fun!!}
d \gtrsim (10^{-7} - 10^{-1}) ~ \textrm{m} .
\end{equation}
There is thus a large possible range of transverse sizes for defects of any dimensionality, which are consistent with the neutron mass variations inside defects that are hinted at by pulsar glitch data.


For defect-induced variation of photon mass, existing astrophysical data and/or further astrophysical observations may provide constraints on the magnitude of the photon mass inside defects. We suggest that limits on the combination of parameters $m_\gamma^2 d$ may be obtained from analysis of short time interval ($t \sim 1$ s $-$ $1$ min) data from non-glitching pulsars. Constraints on related parameters may also be obtained using telescopes, which are sensitive to ranges of different frequencies of electromagnetic radiation, in association with our proposed non-gravitational lensing effects.

\emph{Terrestrial observations.} --- 
We briefly discuss the implications of our findings for the detection of defects using a global network of synchronised atomic clocks, as proposed in \cite{Derevianko2013}. Transition frequencies in atomic and molecular systems are functions of the fundamental constants:
\begin{equation}
\label{omega_FCs}
\frac{\delta \omega (t)}{\omega} = \sum_{X} K_{X} \frac{\delta X(t)}{X} ,
\end{equation}
where the sum is over all relevant fundamental constants $X$, with $K_X$ being the corresponding sensitivity coefficient \cite{Flambaum2006A}. Note that the variation of a dimensionful parameter may be eliminated by a suitable choice of units. Thus for the quantities $X$, we form the dimensionless ratios: $\alpha = e^2/\hbar c$, $m_e/m_p \approx 3 m_e/\Lambda_{\textrm{QCD}}$, $m_q/\Lambda_{\textrm{QCD}}$, and so on. In order to quantify the variation of neutron mass in pulsars due to defects, a suitable dimensionless ratio is $Y = m_n / m_{\textrm{Planck}}$, where $m_{\textrm{Planck}} \propto G^{-1/2}$ is the Planck mass. All atomic and molecular transitions are indepedent of Planck mass. Optical transitions are insensitive to neutron mass variations. For hyperfine and rotational transitions, $K_Y = -1$. For vibrational transitions, $K_Y = -1/2$. A large variety of hyperfine transitions in atomic species \cite{Flambaum2006A,Dinh2009,Berengut2011a,Guena2012}, as well as hyperfine, rotational and vibrational transitions in molecular species \cite{Flambaum2006C,Flambaum2007D,DeMille2008A,Zelevinsky2008A,Kozlov2009A,Kozlov2013B,Kozlov2013C,Flambaum2013B} can be used to search for variations of neutron, proton and electron masses, induced by interaction (\ref{L_Varn_FCs}). 

For a spherical defect with the size of Earth, travelling at a speed of $v \sim 10^{-3} c$, the transit time of a defect passing through a pulsar is $t \sim 40$ s. The required sensitivity of atomic hyperfine and molecular clocks to neutron mass variations hinted by the pulsar glitch data is in the range $\delta m_n/m_n \sim 10^{-11} - 10^{-5}$. Sensitivities of at least the order of $10^{-12}$ on time scales of the order of seconds are achievable with existing Cs ($\sim 10^{-13}$) and Rb ($\sim 10^{-12}$) hyperfine standards, as well as the hydrogen-maser ($\sim 10^{-13}$) \cite{Clocks_comp_2013}, leaving a potentially large range of parameter space that is experimentally accessible with current-generation Earth-based atomic clocks. Global Positioning System satellites already carry on-board Cs and Rb atomic clocks, increasing the range of possibilities for terrestrial-based experiments.

Finally, we mention that alterations in the rotational period of Earth may also be induced by defects passing through Earth. These alterations could be measured by monitoring Earth's angle of rotation over time using an atomic clock. We also mention that the passage of a defect, which consists of axion-like pseudoscalar dark matter, through Earth may give rise to temporary non-zero electric dipole moments (EDMs) for an electron, proton, neutron, nuclei and atoms. For relevant theory, see e.g.~Refs.~\cite{Stadnik2014,Roberts2014}. Such transient EDMs may be sought for with a global network of concurrent EDM experiments. Systems, which may be used for such EDM searches, include a free neutron \cite{Baker2006NEDM}, diamagnetic atoms \cite{Griffith2009improved,Swallows2013,Rosenberry2001atomic,Player1970experiment}, paramagnetic atoms \cite{Ensberg1967,Murthy1989new,Regan2002new},  and molecules \cite{Hudson2011improved,Kara2012measurement,Baron2014,Eckel2013search}.

\emph{Acknowledgements.} --- We would like to thank Julian C.~Berengut, Yvonne Wong, Benjamin M.~Roberts, Alex Kozlov, Andrei Derevianko, Geoff Blewitt, Jeff Sherman and Maxim Pospelov for useful discussions. This work is supported in part by the Australian Research Council and by the Perimeter Institute for Theoretical Physics. Research at the Perimeter Institute is supported by the Government of Canada through Industry Canada and by the Province of Ontario through the Ministry of Economic Development \& Innovation.

\bibliography{TD-DM2}

\begin{thebibliography}{80}%
\makeatletter
\providecommand \@ifxundefined [1]{%
 \@ifx{#1\undefined}
}%
\providecommand \@ifnum [1]{%
 \ifnum #1\expandafter \@firstoftwo
 \else \expandafter \@secondoftwo
 \fi
}%
\providecommand \@ifx [1]{%
 \ifx #1\expandafter \@firstoftwo
 \else \expandafter \@secondoftwo
 \fi
}%
\providecommand \natexlab [1]{#1}%
\providecommand \enquote  [1]{``#1''}%
\providecommand \bibnamefont  [1]{#1}%
\providecommand \bibfnamefont [1]{#1}%
\providecommand \citenamefont [1]{#1}%
\providecommand \href@noop [0]{\@secondoftwo}%
\providecommand \href [0]{\begingroup \@sanitize@url \@href}%
\providecommand \@href[1]{\@@startlink{#1}\@@href}%
\providecommand \@@href[1]{\endgroup#1\@@endlink}%
\providecommand \@sanitize@url [0]{\catcode `\\12\catcode `\$12\catcode
  `\&12\catcode `\#12\catcode `\^12\catcode `\_12\catcode `\%12\relax}%
\providecommand \@@startlink[1]{}%
\providecommand \@@endlink[0]{}%
\providecommand \url  [0]{\begingroup\@sanitize@url \@url }%
\providecommand \@url [1]{\endgroup\@href {#1}{\urlprefix }}%
\providecommand \urlprefix  [0]{URL }%
\providecommand \Eprint [0]{\href }%
\providecommand \doibase [0]{http://dx.doi.org/}%
\providecommand \selectlanguage [0]{\@gobble}%
\providecommand \bibinfo  [0]{\@secondoftwo}%
\providecommand \bibfield  [0]{\@secondoftwo}%
\providecommand \translation [1]{[#1]}%
\providecommand \BibitemOpen [0]{}%
\providecommand \bibitemStop [0]{}%
\providecommand \bibitemNoStop [0]{.\EOS\space}%
\providecommand \EOS [0]{\spacefactor3000\relax}%
\providecommand \BibitemShut  [1]{\csname bibitem#1\endcsname}%
\let\auto@bib@innerbib\@empty
\bibitem [{\citenamefont {Beringer}\ and\ \citenamefont {et~al. {(Particle Data
  Group)}}(2012)}]{Beringer2012PDG}%
  \BibitemOpen
  \bibfield  {author} {\bibinfo {author} {\bibfnamefont {J.}~\bibnamefont
  {Beringer}}\ and\ \bibinfo {author} {\bibnamefont {et~al. {(Particle Data
  Group)}}},\ }\href@noop {} {\bibfield  {journal} {\bibinfo  {journal} {Phys.
  Rev. D}\ }\textbf {\bibinfo {volume} {86}},\ \bibinfo {pages} {010001}
  (\bibinfo {year} {2012})}\BibitemShut {NoStop}%
\bibitem [{\citenamefont {Vilenkin}(1985)}]{Vilenkin1985}%
  \BibitemOpen
  \bibfield  {author} {\bibinfo {author} {\bibfnamefont {A.}~\bibnamefont
  {Vilenkin}},\ }\href@noop {} {\bibfield  {journal} {\bibinfo  {journal}
  {Phys. Rep.}\ }\textbf {\bibinfo {volume} {121}},\ \bibinfo {pages} {263}
  (\bibinfo {year} {1985})}\BibitemShut {NoStop}%
\bibitem [{\citenamefont {Vilenkin}\ and\ \citenamefont
  {Shellard}(1994)}]{TDDM_Book1}%
  \BibitemOpen
  \bibfield  {author} {\bibinfo {author} {\bibfnamefont {A.}~\bibnamefont
  {Vilenkin}}\ and\ \bibinfo {author} {\bibfnamefont {E.}~\bibnamefont
  {Shellard}},\ }\href@noop {} {\emph {\bibinfo {title} {{Cosmic Strings and
  Other Topological Defects}}}}\ (\bibinfo  {publisher} {Cambridge University
  Press},\ \bibinfo {address} {Cambridge},\ \bibinfo {year} {1994})\BibitemShut
  {NoStop}%
\bibitem [{\citenamefont {Schneider}\ \emph {et~al.}(1999)\citenamefont
  {Schneider}, \citenamefont {Ehlers},\ and\ \citenamefont
  {Falco}}]{TDDM_Book2}%
  \BibitemOpen
  \bibfield  {author} {\bibinfo {author} {\bibfnamefont {P.}~\bibnamefont
  {Schneider}}, \bibinfo {author} {\bibfnamefont {J.}~\bibnamefont {Ehlers}}, \
  and\ \bibinfo {author} {\bibfnamefont {E.~E.}\ \bibnamefont {Falco}},\
  }\href@noop {} {\emph {\bibinfo {title} {{Gravitational Lenses}}}}\ (\bibinfo
   {publisher} {Springer-Verlag},\ \bibinfo {address} {Berlin Heidelberg},\
  \bibinfo {year} {1999})\BibitemShut {NoStop}%
\bibitem [{\citenamefont {Cline}(2001)}]{TDDM_Book3}%
  \BibitemOpen
  \bibinfo {editor} {\bibfnamefont {D.~B.}\ \bibnamefont {Cline}},\ ed.,\
  \href@noop {} {\emph {\bibinfo {title} {{Sources and Detection of Dark Matter
  and Dark Energy in the Universe: Fourth International Symposium Held at
  Marina del Rey, CA, USA February 23-25, 2000}}}}\ (\bibinfo  {publisher}
  {Springer-Verlag},\ \bibinfo {address} {Berlin Heidelberg},\ \bibinfo {year}
  {2001})\BibitemShut {NoStop}%
\bibitem [{\citenamefont {Ade}\ and\ \citenamefont
  {Others}(2013)}]{Planck2013-Defects}%
  \BibitemOpen
  \bibfield  {author} {\bibinfo {author} {\bibfnamefont {P.~A.~R.}\
  \bibnamefont {Ade}}\ and\ \bibinfo {author} {\bibnamefont {Others}},\
  }\href@noop {} {\bibfield  {journal} {\bibinfo  {journal} {arXiv:1303.5085}\
  } (\bibinfo {year} {2013})}\BibitemShut {NoStop}%
\bibitem [{\citenamefont {Ade}\ and\ \citenamefont {Others}(2014)}]{BICEP2014}%
  \BibitemOpen
  \bibfield  {author} {\bibinfo {author} {\bibfnamefont {P.~A.~R.}\
  \bibnamefont {Ade}}\ and\ \bibinfo {author} {\bibnamefont {Others}},\
  }\href@noop {} {\bibfield  {journal} {\bibinfo  {journal} {Phys. Rev. Lett.}\
  }\textbf {\bibinfo {volume} {112}},\ \bibinfo {pages} {241101} (\bibinfo
  {year} {2014})}\BibitemShut {NoStop}%
\bibitem [{\citenamefont {Moss}(2014)}]{Moss2014-BICEP2}%
  \BibitemOpen
  \bibfield  {author} {\bibinfo {author} {\bibfnamefont {L.}~\bibnamefont
  {Moss}, \bibfnamefont {A.~Pogosian}},\ }\href@noop {} {\bibfield  {journal}
  {\bibinfo  {journal} {Phys. Rev. Lett.}\ }\textbf {\bibinfo {volume} {112}},\
  \bibinfo {pages} {171302} (\bibinfo {year} {2014})}\BibitemShut {NoStop}%
\bibitem [{\citenamefont {Pospelov}\ \emph {et~al.}(2013)\citenamefont
  {Pospelov}, \citenamefont {Pustelny}, \citenamefont {Ledbetter},
  \citenamefont {Kimball}, \citenamefont {Gawlik},\ and\ \citenamefont
  {Budker}}]{Pospelov2013}%
  \BibitemOpen
  \bibfield  {author} {\bibinfo {author} {\bibfnamefont {M.}~\bibnamefont
  {Pospelov}}, \bibinfo {author} {\bibfnamefont {S.}~\bibnamefont {Pustelny}},
  \bibinfo {author} {\bibfnamefont {M.~P.}\ \bibnamefont {Ledbetter}}, \bibinfo
  {author} {\bibfnamefont {D.~F.~J.}\ \bibnamefont {Kimball}}, \bibinfo
  {author} {\bibfnamefont {W.}~\bibnamefont {Gawlik}}, \ and\ \bibinfo {author}
  {\bibfnamefont {D.}~\bibnamefont {Budker}},\ }\href {\doibase
  10.1103/PhysRevLett.110.021803} {\bibfield  {journal} {\bibinfo  {journal}
  {Phys. Rev. Lett.}\ }\textbf {\bibinfo {volume} {110}},\ \bibinfo {pages}
  {021803} (\bibinfo {year} {2013})}\BibitemShut {NoStop}%
\bibitem [{\citenamefont {Derevianko}\ and\ \citenamefont
  {Pospelov}(2013)}]{Derevianko2013}%
  \BibitemOpen
  \bibfield  {author} {\bibinfo {author} {\bibfnamefont {A.}~\bibnamefont
  {Derevianko}}\ and\ \bibinfo {author} {\bibfnamefont {M.}~\bibnamefont
  {Pospelov}},\ }\href@noop {} {\bibfield  {journal} {\bibinfo  {journal}
  {arXiv:1311.1244}\ } (\bibinfo {year} {2013})}\BibitemShut {NoStop}%
\bibitem [{\citenamefont {Stairs}(2003)}]{Stairs2003Alive}%
  \BibitemOpen
  \bibfield  {author} {\bibinfo {author} {\bibfnamefont {I.~H.}\ \bibnamefont
  {Stairs}},\ }\href {http://www.livingreviews.org/lrr-2003-5} {\bibfield
  {journal} {\bibinfo  {journal} {Living Rev. Relativ.}\ }\textbf {\bibinfo
  {volume} {6}},\ \bibinfo {pages} {5} (\bibinfo {year} {2003})}\BibitemShut
  {NoStop}%
\bibitem [{\citenamefont {Lyne}\ and\ \citenamefont
  {Graham-Smith}(2006)}]{Lyne2012Pulsar}%
  \BibitemOpen
  \bibfield  {author} {\bibinfo {author} {\bibfnamefont {A.~G.}\ \bibnamefont
  {Lyne}}\ and\ \bibinfo {author} {\bibfnamefont {F.}~\bibnamefont
  {Graham-Smith}},\ }\href@noop {} {\emph {\bibinfo {title} {{Pulsar
  Astronomy}}}},\ \bibinfo {edition} {3rd}\ ed.\ (\bibinfo  {publisher}
  {Cambridge University Press},\ \bibinfo {address} {Cambridge},\ \bibinfo
  {year} {2006})\BibitemShut {NoStop}%
\bibitem [{\citenamefont {Hinkley}\ \emph {et~al.}(2013)\citenamefont
  {Hinkley}, \citenamefont {Sherman}, \citenamefont {Phillips}, \citenamefont
  {Schioppo}, \citenamefont {Lemke}, \citenamefont {Beloy}, \citenamefont
  {Pizzocaro}, \citenamefont {Oates},\ and\ \citenamefont
  {Ludlow}}]{Hinkley2013}%
  \BibitemOpen
  \bibfield  {author} {\bibinfo {author} {\bibfnamefont {N.}~\bibnamefont
  {Hinkley}}, \bibinfo {author} {\bibfnamefont {J.~A.}\ \bibnamefont
  {Sherman}}, \bibinfo {author} {\bibfnamefont {N.~B.}\ \bibnamefont
  {Phillips}}, \bibinfo {author} {\bibfnamefont {M.}~\bibnamefont {Schioppo}},
  \bibinfo {author} {\bibfnamefont {N.~D.}\ \bibnamefont {Lemke}}, \bibinfo
  {author} {\bibfnamefont {K.}~\bibnamefont {Beloy}}, \bibinfo {author}
  {\bibfnamefont {M.}~\bibnamefont {Pizzocaro}}, \bibinfo {author}
  {\bibfnamefont {C.~W.}\ \bibnamefont {Oates}}, \ and\ \bibinfo {author}
  {\bibfnamefont {A.~D.}\ \bibnamefont {Ludlow}},\ }\href@noop {} {\bibfield
  {journal} {\bibinfo  {journal} {Science}\ }\textbf {\bibinfo {volume}
  {341}},\ \bibinfo {pages} {1215} (\bibinfo {year} {2013})}\BibitemShut
  {NoStop}%
\bibitem [{\citenamefont {Bloom}\ \emph {et~al.}(2014)\citenamefont {Bloom},
  \citenamefont {Nicholson}, \citenamefont {Williams}, \citenamefont
  {Campbell}, \citenamefont {Bishof}, \citenamefont {Zhang}, \citenamefont
  {Zhang}, \citenamefont {Bromley},\ and\ \citenamefont {Ye}}]{Bloom2014}%
  \BibitemOpen
  \bibfield  {author} {\bibinfo {author} {\bibfnamefont {B.~J.}\ \bibnamefont
  {Bloom}}, \bibinfo {author} {\bibfnamefont {T.~L.}\ \bibnamefont
  {Nicholson}}, \bibinfo {author} {\bibfnamefont {J.~R.}\ \bibnamefont
  {Williams}}, \bibinfo {author} {\bibfnamefont {S.~L.}\ \bibnamefont
  {Campbell}}, \bibinfo {author} {\bibfnamefont {M.}~\bibnamefont {Bishof}},
  \bibinfo {author} {\bibfnamefont {X.}~\bibnamefont {Zhang}}, \bibinfo
  {author} {\bibfnamefont {W.}~\bibnamefont {Zhang}}, \bibinfo {author}
  {\bibfnamefont {S.~L.}\ \bibnamefont {Bromley}}, \ and\ \bibinfo {author}
  {\bibfnamefont {J.}~\bibnamefont {Ye}},\ }\href@noop {} {\bibfield  {journal}
  {\bibinfo  {journal} {Nature}\ }\textbf {\bibinfo {volume} {506}},\ \bibinfo
  {pages} {71} (\bibinfo {year} {2014})}\BibitemShut {NoStop}%
\bibitem [{\citenamefont {Baym}\ \emph {et~al.}(1969)\citenamefont {Baym},
  \citenamefont {Pethick},\ and\ \citenamefont {Pines}}]{Baym1969}%
  \BibitemOpen
  \bibfield  {author} {\bibinfo {author} {\bibfnamefont {G.}~\bibnamefont
  {Baym}}, \bibinfo {author} {\bibfnamefont {C.}~\bibnamefont {Pethick}}, \
  and\ \bibinfo {author} {\bibfnamefont {D.}~\bibnamefont {Pines}},\
  }\href@noop {} {\bibfield  {journal} {\bibinfo  {journal} {Nature}\ }\textbf
  {\bibinfo {volume} {224}},\ \bibinfo {pages} {872} (\bibinfo {year}
  {1969})}\BibitemShut {NoStop}%
\bibitem [{\citenamefont {Anderson}\ and\ \citenamefont
  {Itoh}(1975)}]{Anderson1975}%
  \BibitemOpen
  \bibfield  {author} {\bibinfo {author} {\bibfnamefont {P.~W.}\ \bibnamefont
  {Anderson}}\ and\ \bibinfo {author} {\bibfnamefont {N.}~\bibnamefont
  {Itoh}},\ }\href@noop {} {\bibfield  {journal} {\bibinfo  {journal} {Nature}\
  }\textbf {\bibinfo {volume} {256}},\ \bibinfo {pages} {25} (\bibinfo {year}
  {1975})}\BibitemShut {NoStop}%
\bibitem [{\citenamefont {Alpar}\ \emph {et~al.}(1984)\citenamefont {Alpar},
  \citenamefont {Pines}, \citenamefont {Anderson},\ and\ \citenamefont
  {Shaham}}]{Alpar1984}%
  \BibitemOpen
  \bibfield  {author} {\bibinfo {author} {\bibfnamefont {M.~A.}\ \bibnamefont
  {Alpar}}, \bibinfo {author} {\bibfnamefont {D.}~\bibnamefont {Pines}},
  \bibinfo {author} {\bibfnamefont {P.~W.}\ \bibnamefont {Anderson}}, \ and\
  \bibinfo {author} {\bibfnamefont {J.}~\bibnamefont {Shaham}},\ }\href@noop {}
  {\bibfield  {journal} {\bibinfo  {journal} {Astrophys. J.}\ }\textbf
  {\bibinfo {volume} {276}},\ \bibinfo {pages} {325} (\bibinfo {year}
  {1984})}\BibitemShut {NoStop}%
\bibitem [{\citenamefont {Pines}\ and\ \citenamefont
  {Alpar}(1985)}]{Pines1985}%
  \BibitemOpen
  \bibfield  {author} {\bibinfo {author} {\bibfnamefont {D.}~\bibnamefont
  {Pines}}\ and\ \bibinfo {author} {\bibfnamefont {M.~A.}\ \bibnamefont
  {Alpar}},\ }\href@noop {} {\bibfield  {journal} {\bibinfo  {journal}
  {Nature}\ }\textbf {\bibinfo {volume} {316}},\ \bibinfo {pages} {27}
  (\bibinfo {year} {1985})}\BibitemShut {NoStop}%
\bibitem [{\citenamefont {Takatsuka}\ and\ \citenamefont
  {Tamagaki}(1988)}]{Takatsuka1988}%
  \BibitemOpen
  \bibfield  {author} {\bibinfo {author} {\bibfnamefont {T.}~\bibnamefont
  {Takatsuka}}\ and\ \bibinfo {author} {\bibfnamefont {R.}~\bibnamefont
  {Tamagaki}},\ }\href@noop {} {\bibfield  {journal} {\bibinfo  {journal}
  {Nucl. Phys. A}\ }\textbf {\bibinfo {volume} {478}},\ \bibinfo {pages} {785}
  (\bibinfo {year} {1988})}\BibitemShut {NoStop}%
\bibitem [{\citenamefont {Ravenhall}\ and\ \citenamefont
  {Pethick}(1994)}]{Ravenhall1994}%
  \BibitemOpen
  \bibfield  {author} {\bibinfo {author} {\bibfnamefont {D.~G.}\ \bibnamefont
  {Ravenhall}}\ and\ \bibinfo {author} {\bibfnamefont {C.~J.}\ \bibnamefont
  {Pethick}},\ }\href@noop {} {\bibfield  {journal} {\bibinfo  {journal}
  {Astrophys. J.}\ }\textbf {\bibinfo {volume} {424}},\ \bibinfo {pages} {846}
  (\bibinfo {year} {1994})}\BibitemShut {NoStop}%
\bibitem [{\citenamefont {Link}\ \emph {et~al.}(1999)\citenamefont {Link},
  \citenamefont {Epstein},\ and\ \citenamefont {Lattimer}}]{Link1999}%
  \BibitemOpen
  \bibfield  {author} {\bibinfo {author} {\bibfnamefont {B.}~\bibnamefont
  {Link}}, \bibinfo {author} {\bibfnamefont {R.~I.}\ \bibnamefont {Epstein}}, \
  and\ \bibinfo {author} {\bibfnamefont {J.~M.}\ \bibnamefont {Lattimer}},\
  }\href@noop {} {\bibfield  {journal} {\bibinfo  {journal} {Phys. Rev. Lett.}\
  }\textbf {\bibinfo {volume} {83}},\ \bibinfo {pages} {3362} (\bibinfo {year}
  {1999})}\BibitemShut {NoStop}%
\bibitem [{\citenamefont {Andersson}\ \emph {et~al.}(2003)\citenamefont
  {Andersson}, \citenamefont {Comer},\ and\ \citenamefont
  {Prix}}]{Andersson2003}%
  \BibitemOpen
  \bibfield  {author} {\bibinfo {author} {\bibfnamefont {N.}~\bibnamefont
  {Andersson}}, \bibinfo {author} {\bibfnamefont {G.~L.}\ \bibnamefont
  {Comer}}, \ and\ \bibinfo {author} {\bibfnamefont {R.}~\bibnamefont {Prix}},\
  }\href@noop {} {\bibfield  {journal} {\bibinfo  {journal} {Phys. Rev. Lett.}\
  }\textbf {\bibinfo {volume} {90}},\ \bibinfo {pages} {091101} (\bibinfo
  {year} {2003})}\BibitemShut {NoStop}%
\bibitem [{\citenamefont {Chamel}(2005)}]{Chamel2005}%
  \BibitemOpen
  \bibfield  {author} {\bibinfo {author} {\bibfnamefont {N.}~\bibnamefont
  {Chamel}},\ }\href@noop {} {\bibfield  {journal} {\bibinfo  {journal} {Nucl.
  Phys. A}\ }\textbf {\bibinfo {volume} {747}},\ \bibinfo {pages} {109}
  (\bibinfo {year} {2005})}\BibitemShut {NoStop}%
\bibitem [{\citenamefont {Donati}\ and\ \citenamefont
  {Pizzochero}(2006)}]{Donati2006}%
  \BibitemOpen
  \bibfield  {author} {\bibinfo {author} {\bibfnamefont {P.}~\bibnamefont
  {Donati}}\ and\ \bibinfo {author} {\bibfnamefont {P.~M.}\ \bibnamefont
  {Pizzochero}},\ }\href@noop {} {\bibfield  {journal} {\bibinfo  {journal}
  {Phys. Lett. B}\ }\textbf {\bibinfo {volume} {640}},\ \bibinfo {pages} {74}
  (\bibinfo {year} {2006})}\BibitemShut {NoStop}%
\bibitem [{\citenamefont {Melatos}\ and\ \citenamefont
  {Peralta}(2007)}]{Melatos2007}%
  \BibitemOpen
  \bibfield  {author} {\bibinfo {author} {\bibfnamefont {A.}~\bibnamefont
  {Melatos}}\ and\ \bibinfo {author} {\bibfnamefont {C.}~\bibnamefont
  {Peralta}},\ }\href@noop {} {\bibfield  {journal} {\bibinfo  {journal}
  {Astrophys. J. Lett.}\ }\textbf {\bibinfo {volume} {662}},\ \bibinfo {pages}
  {L99} (\bibinfo {year} {2007})}\BibitemShut {NoStop}%
\bibitem [{\citenamefont {Glampedakis}\ and\ \citenamefont
  {Andersson}(2009)}]{Glampedakis2009}%
  \BibitemOpen
  \bibfield  {author} {\bibinfo {author} {\bibfnamefont {K.}~\bibnamefont
  {Glampedakis}}\ and\ \bibinfo {author} {\bibfnamefont {N.}~\bibnamefont
  {Andersson}},\ }\href@noop {} {\bibfield  {journal} {\bibinfo  {journal}
  {Phys. Rev. Lett.}\ }\textbf {\bibinfo {volume} {102}},\ \bibinfo {pages}
  {141101} (\bibinfo {year} {2009})}\BibitemShut {NoStop}%
\bibitem [{\citenamefont {{Van Eysden}}\ and\ \citenamefont
  {Melatos}(2010)}]{Van2010pulsar}%
  \BibitemOpen
  \bibfield  {author} {\bibinfo {author} {\bibfnamefont {C.~A.}\ \bibnamefont
  {{Van Eysden}}}\ and\ \bibinfo {author} {\bibfnamefont {A.}~\bibnamefont
  {Melatos}},\ }\href@noop {} {\bibfield  {journal} {\bibinfo  {journal} {Mon.
  Not. R. Astron. Soc.}\ }\textbf {\bibinfo {volume} {409}},\ \bibinfo {pages}
  {1253} (\bibinfo {year} {2010})}\BibitemShut {NoStop}%
\bibitem [{\citenamefont {Pizzochero}(2011)}]{Pizzochero2011}%
  \BibitemOpen
  \bibfield  {author} {\bibinfo {author} {\bibfnamefont {P.~M.}\ \bibnamefont
  {Pizzochero}},\ }\href@noop {} {\bibfield  {journal} {\bibinfo  {journal}
  {Astrophys. J. Lett.}\ }\textbf {\bibinfo {volume} {743}},\ \bibinfo {pages}
  {L20} (\bibinfo {year} {2011})}\BibitemShut {NoStop}%
\bibitem [{\citenamefont {Andersson}\ \emph {et~al.}(2012)\citenamefont
  {Andersson}, \citenamefont {Glampedakis}, \citenamefont {Ho},\ and\
  \citenamefont {Espinoza}}]{Andersson2012pulsar}%
  \BibitemOpen
  \bibfield  {author} {\bibinfo {author} {\bibfnamefont {N.}~\bibnamefont
  {Andersson}}, \bibinfo {author} {\bibfnamefont {K.}~\bibnamefont
  {Glampedakis}}, \bibinfo {author} {\bibfnamefont {W.~C.~G.}\ \bibnamefont
  {Ho}}, \ and\ \bibinfo {author} {\bibfnamefont {C.~M.}\ \bibnamefont
  {Espinoza}},\ }\href@noop {} {\bibfield  {journal} {\bibinfo  {journal}
  {Phys. Rev. Lett.}\ }\textbf {\bibinfo {volume} {109}},\ \bibinfo {pages}
  {241103} (\bibinfo {year} {2012})}\BibitemShut {NoStop}%
\bibitem [{\citenamefont {Chamel}(2012)}]{Chamel2012}%
  \BibitemOpen
  \bibfield  {author} {\bibinfo {author} {\bibfnamefont {N.}~\bibnamefont
  {Chamel}},\ }\href@noop {} {\bibfield  {journal} {\bibinfo  {journal} {Phys.
  Rev. C}\ }\textbf {\bibinfo {volume} {85}},\ \bibinfo {pages} {035801}
  (\bibinfo {year} {2012})}\BibitemShut {NoStop}%
\bibitem [{\citenamefont {Steiner}\ \emph {et~al.}(2014)\citenamefont
  {Steiner}, \citenamefont {Gandolfi}, \citenamefont {Fattoyev},\ and\
  \citenamefont {Newton}}]{Steiner2014glitchcrust}%
  \BibitemOpen
  \bibfield  {author} {\bibinfo {author} {\bibfnamefont {A.~W.}\ \bibnamefont
  {Steiner}}, \bibinfo {author} {\bibfnamefont {S.}~\bibnamefont {Gandolfi}},
  \bibinfo {author} {\bibfnamefont {F.~J.}\ \bibnamefont {Fattoyev}}, \ and\
  \bibinfo {author} {\bibfnamefont {W.~G.}\ \bibnamefont {Newton}},\
  }\href@noop {} {\bibfield  {journal} {\bibinfo  {journal} {arXiv:1403.7546}\
  } (\bibinfo {year} {2014})}\BibitemShut {NoStop}%
\bibitem [{\citenamefont {Piekarewicz}\ \emph {et~al.}(2014)\citenamefont
  {Piekarewicz}, \citenamefont {Fattoyev},\ and\ \citenamefont
  {Horowitz}}]{Piekarewicz2014}%
  \BibitemOpen
  \bibfield  {author} {\bibinfo {author} {\bibfnamefont {J.}~\bibnamefont
  {Piekarewicz}}, \bibinfo {author} {\bibfnamefont {F.~J.}\ \bibnamefont
  {Fattoyev}}, \ and\ \bibinfo {author} {\bibfnamefont {C.~J.}\ \bibnamefont
  {Horowitz}},\ }\href@noop {} {\bibfield  {journal} {\bibinfo  {journal}
  {arXiv:1404.2660}\ } (\bibinfo {year} {2014})}\BibitemShut {NoStop}%
\bibitem [{\citenamefont {Li}\ \emph {et~al.}(2014)\citenamefont {Li},
  \citenamefont {Wang}, \citenamefont {Shao},\ and\ \citenamefont
  {Xu}}]{Li2014glitches}%
  \BibitemOpen
  \bibfield  {author} {\bibinfo {author} {\bibfnamefont {A.}~\bibnamefont
  {Li}}, \bibinfo {author} {\bibfnamefont {J.}~\bibnamefont {Wang}}, \bibinfo
  {author} {\bibfnamefont {L.}~\bibnamefont {Shao}}, \ and\ \bibinfo {author}
  {\bibfnamefont {R.-X.}\ \bibnamefont {Xu}},\ }\href@noop {} {\bibfield
  {journal} {\bibinfo  {journal} {arXiv:1406.4994}\ } (\bibinfo {year}
  {2014})}\BibitemShut {NoStop}%
\bibitem [{\citenamefont {CSIRO}(2014)}]{Pulsar_CSIRO_Database}%
  \BibitemOpen
  \bibfield  {author} {\bibinfo {author} {\bibnamefont {CSIRO}},\ }\href
  {http://www.atnf.csiro.au/people/pulsar/psrcat/glitchTbl.html} {\enquote
  {\bibinfo {title} {{ATNF Pulsar Catalogue: Glitch Parameters,
  http://www.atnf.csiro.au/people/pulsar/psrcat/glitchTbl.html}},}\ } (\bibinfo
  {year} {2014})\BibitemShut {NoStop}%
\bibitem [{\citenamefont {Olive}\ and\ \citenamefont
  {Pospelov}(2008)}]{Olive2008P}%
  \BibitemOpen
  \bibfield  {author} {\bibinfo {author} {\bibfnamefont {K.~A.}\ \bibnamefont
  {Olive}}\ and\ \bibinfo {author} {\bibfnamefont {M.}~\bibnamefont
  {Pospelov}},\ }\href@noop {} {\bibfield  {journal} {\bibinfo  {journal}
  {Phys. Rev. D}\ }\textbf {\bibinfo {volume} {77}},\ \bibinfo {pages} {043524}
  (\bibinfo {year} {2008})}\BibitemShut {NoStop}%
\bibitem [{\citenamefont {Brown}(2000)}]{Brown2000neutron}%
  \BibitemOpen
  \bibfield  {author} {\bibinfo {author} {\bibfnamefont {B.~A.}\ \bibnamefont
  {Brown}},\ }\href@noop {} {\bibfield  {journal} {\bibinfo  {journal} {Phys.
  Rev. Lett.}\ }\textbf {\bibinfo {volume} {85}},\ \bibinfo {pages} {5296}
  (\bibinfo {year} {2000})}\BibitemShut {NoStop}%
\bibitem [{\citenamefont {Typel}\ and\ \citenamefont
  {Brown}(2001)}]{Typel2001neutron}%
  \BibitemOpen
  \bibfield  {author} {\bibinfo {author} {\bibfnamefont {S.}~\bibnamefont
  {Typel}}\ and\ \bibinfo {author} {\bibfnamefont {B.~A.}\ \bibnamefont
  {Brown}},\ }\href@noop {} {\bibfield  {journal} {\bibinfo  {journal} {Phys.
  Rev. C}\ }\textbf {\bibinfo {volume} {64}},\ \bibinfo {pages} {027302}
  (\bibinfo {year} {2001})}\BibitemShut {NoStop}%
\bibitem [{\citenamefont {Steiner}\ \emph {et~al.}(2005)\citenamefont
  {Steiner}, \citenamefont {Prakash}, \citenamefont {Lattimer},\ and\
  \citenamefont {Ellis}}]{Steiner2005isospin}%
  \BibitemOpen
  \bibfield  {author} {\bibinfo {author} {\bibfnamefont {A.~W.}\ \bibnamefont
  {Steiner}}, \bibinfo {author} {\bibfnamefont {M.}~\bibnamefont {Prakash}},
  \bibinfo {author} {\bibfnamefont {J.~M.}\ \bibnamefont {Lattimer}}, \ and\
  \bibinfo {author} {\bibfnamefont {P.~J.}\ \bibnamefont {Ellis}},\ }\href@noop
  {} {\bibfield  {journal} {\bibinfo  {journal} {Phys. Rep.}\ }\textbf
  {\bibinfo {volume} {411}},\ \bibinfo {pages} {325} (\bibinfo {year}
  {2005})}\BibitemShut {NoStop}%
\bibitem [{\citenamefont {Steiner}\ \emph {et~al.}(2013)\citenamefont
  {Steiner}, \citenamefont {Lattimer},\ and\ \citenamefont
  {Brown}}]{Steiner2013neutron}%
  \BibitemOpen
  \bibfield  {author} {\bibinfo {author} {\bibfnamefont {A.~W.}\ \bibnamefont
  {Steiner}}, \bibinfo {author} {\bibfnamefont {J.~M.}\ \bibnamefont
  {Lattimer}}, \ and\ \bibinfo {author} {\bibfnamefont {E.~F.}\ \bibnamefont
  {Brown}},\ }\href@noop {} {\bibfield  {journal} {\bibinfo  {journal}
  {Astrophys. J. Lett.}\ }\textbf {\bibinfo {volume} {765}},\ \bibinfo {pages}
  {L5} (\bibinfo {year} {2013})}\BibitemShut {NoStop}%
\bibitem [{\citenamefont {Brown}(2013)}]{Brown2013constraints}%
  \BibitemOpen
  \bibfield  {author} {\bibinfo {author} {\bibfnamefont {B.~A.}\ \bibnamefont
  {Brown}},\ }\href@noop {} {\bibfield  {journal} {\bibinfo  {journal} {Phys.
  Rev. Lett.}\ }\textbf {\bibinfo {volume} {111}},\ \bibinfo {pages} {232502}
  (\bibinfo {year} {2013})}\BibitemShut {NoStop}%
\bibitem [{\citenamefont {Brown}\ and\ \citenamefont
  {Schwenk}(2014)}]{Brown2014constraints}%
  \BibitemOpen
  \bibfield  {author} {\bibinfo {author} {\bibfnamefont {B.~A.}\ \bibnamefont
  {Brown}}\ and\ \bibinfo {author} {\bibfnamefont {A.}~\bibnamefont
  {Schwenk}},\ }\href@noop {} {\bibfield  {journal} {\bibinfo  {journal} {Phys.
  Rev. C}\ }\textbf {\bibinfo {volume} {89}},\ \bibinfo {pages} {011307}
  (\bibinfo {year} {2014})}\BibitemShut {NoStop}%
\bibitem [{\citenamefont {Landau}\ and\ \citenamefont {Lifshitz}(1980)}]{LL5}%
  \BibitemOpen
  \bibfield  {author} {\bibinfo {author} {\bibfnamefont {L.~D.}\ \bibnamefont
  {Landau}}\ and\ \bibinfo {author} {\bibfnamefont {E.~M.}\ \bibnamefont
  {Lifshitz}},\ }\href@noop {} {\emph {\bibinfo {title} {{Statistical Physics,
  Part 1}}}},\ \bibinfo {edition} {3rd}\ ed.\ (\bibinfo  {publisher}
  {Butterworth-Heinemann},\ \bibinfo {address} {Oxford},\ \bibinfo {year}
  {1980})\BibitemShut {NoStop}%
\bibitem [{\citenamefont {Allan}\ \emph {et~al.}(1997)\citenamefont {Allan},
  \citenamefont {Ashby},\ and\ \citenamefont {Hodge}}]{Allan1997}%
  \BibitemOpen
  \bibfield  {author} {\bibinfo {author} {\bibfnamefont {D.~W.}\ \bibnamefont
  {Allan}}, \bibinfo {author} {\bibfnamefont {N.}~\bibnamefont {Ashby}}, \ and\
  \bibinfo {author} {\bibfnamefont {C.~C.}\ \bibnamefont {Hodge}},\ }\href@noop
  {} {\emph {\bibinfo {title} {{The science of timekeeping}}}},\ \bibinfo
  {number} {1289}\ (\bibinfo  {publisher} {Hewlett-Packard},\ \bibinfo {year}
  {1997})\BibitemShut {NoStop}%
\bibitem [{\citenamefont {Peik}\ \emph {et~al.}(2004)\citenamefont {Peik},
  \citenamefont {Lipphardt}, \citenamefont {Schnatz}, \citenamefont
  {Schneider}, \citenamefont {Tamm},\ and\ \citenamefont
  {Karshenboim}}]{Peik2004}%
  \BibitemOpen
  \bibfield  {author} {\bibinfo {author} {\bibfnamefont {E.}~\bibnamefont
  {Peik}}, \bibinfo {author} {\bibfnamefont {B.}~\bibnamefont {Lipphardt}},
  \bibinfo {author} {\bibfnamefont {H.}~\bibnamefont {Schnatz}}, \bibinfo
  {author} {\bibfnamefont {T.}~\bibnamefont {Schneider}}, \bibinfo {author}
  {\bibfnamefont {C.}~\bibnamefont {Tamm}}, \ and\ \bibinfo {author}
  {\bibfnamefont {S.~G.}\ \bibnamefont {Karshenboim}},\ }\href {\doibase
  10.1103/PhysRevLett.93.170801} {\bibfield  {journal} {\bibinfo  {journal}
  {Phys. Rev. Lett.}\ }\textbf {\bibinfo {volume} {93}},\ \bibinfo {pages}
  {170801} (\bibinfo {year} {2004})}\BibitemShut {NoStop}%
\bibitem [{\citenamefont {Takamoto}\ \emph {et~al.}(2005)\citenamefont
  {Takamoto}, \citenamefont {Hong}, \citenamefont {Higashi},\ and\
  \citenamefont {Katori}}]{Takamoto2005opticalSrLAttice}%
  \BibitemOpen
  \bibfield  {author} {\bibinfo {author} {\bibfnamefont {M.}~\bibnamefont
  {Takamoto}}, \bibinfo {author} {\bibfnamefont {F.-L.}\ \bibnamefont {Hong}},
  \bibinfo {author} {\bibfnamefont {R.}~\bibnamefont {Higashi}}, \ and\
  \bibinfo {author} {\bibfnamefont {H.}~\bibnamefont {Katori}},\ }\href@noop {}
  {\bibfield  {journal} {\bibinfo  {journal} {Nature}\ }\textbf {\bibinfo
  {volume} {435}},\ \bibinfo {pages} {321} (\bibinfo {year}
  {2005})}\BibitemShut {NoStop}%
\bibitem [{\citenamefont {Oskay}\ \emph {et~al.}(2006)\citenamefont {Oskay},
  \citenamefont {Diddams}, \citenamefont {Donley}, \citenamefont {Fortier},
  \citenamefont {Heavner}, \citenamefont {Hollberg}, \citenamefont {Itano},
  \citenamefont {Jefferts}, \citenamefont {Delaney}, \citenamefont {Kim},\ and\
  \citenamefont {Others}}]{Oskay2006}%
  \BibitemOpen
  \bibfield  {author} {\bibinfo {author} {\bibfnamefont {W.~H.}\ \bibnamefont
  {Oskay}}, \bibinfo {author} {\bibfnamefont {S.~A.}\ \bibnamefont {Diddams}},
  \bibinfo {author} {\bibfnamefont {E.~A.}\ \bibnamefont {Donley}}, \bibinfo
  {author} {\bibfnamefont {T.~M.}\ \bibnamefont {Fortier}}, \bibinfo {author}
  {\bibfnamefont {T.~P.}\ \bibnamefont {Heavner}}, \bibinfo {author}
  {\bibfnamefont {L.}~\bibnamefont {Hollberg}}, \bibinfo {author}
  {\bibfnamefont {W.~M.}\ \bibnamefont {Itano}}, \bibinfo {author}
  {\bibfnamefont {S.~R.}\ \bibnamefont {Jefferts}}, \bibinfo {author}
  {\bibfnamefont {M.~J.}\ \bibnamefont {Delaney}}, \bibinfo {author}
  {\bibfnamefont {K.}~\bibnamefont {Kim}}, \ and\ \bibinfo {author}
  {\bibnamefont {Others}},\ }\href@noop {} {\bibfield  {journal} {\bibinfo
  {journal} {Phys. Rev. Lett.}\ }\textbf {\bibinfo {volume} {97}},\ \bibinfo
  {pages} {020801} (\bibinfo {year} {2006})}\BibitemShut {NoStop}%
\bibitem [{\citenamefont {Hartnett}\ \emph {et~al.}(2006)\citenamefont
  {Hartnett}, \citenamefont {Locke}, \citenamefont {Ivanov}, \citenamefont
  {Tobar},\ and\ \citenamefont {Stanwix}}]{Tobar2006Sapphire}%
  \BibitemOpen
  \bibfield  {author} {\bibinfo {author} {\bibfnamefont {J.~G.}\ \bibnamefont
  {Hartnett}}, \bibinfo {author} {\bibfnamefont {C.~R.}\ \bibnamefont {Locke}},
  \bibinfo {author} {\bibfnamefont {E.~N.}\ \bibnamefont {Ivanov}}, \bibinfo
  {author} {\bibfnamefont {M.~E.}\ \bibnamefont {Tobar}}, \ and\ \bibinfo
  {author} {\bibfnamefont {P.~L.}\ \bibnamefont {Stanwix}},\ }\href@noop {}
  {\bibfield  {journal} {\bibinfo  {journal} {Appl. Phys. Lett.}\ }\textbf
  {\bibinfo {volume} {89}},\ \bibinfo {pages} {203513} (\bibinfo {year}
  {2006})}\BibitemShut {NoStop}%
\bibitem [{\citenamefont {Ludlow}\ \emph {et~al.}(2008)\citenamefont {Ludlow},
  \citenamefont {Zelevinsky}, \citenamefont {Campbell}, \citenamefont {Blatt},
  \citenamefont {Boyd}, \citenamefont {{De Miranda}}, \citenamefont {Martin},
  \citenamefont {Thomsen}, \citenamefont {Foreman}, \citenamefont {Ye},\ and\
  \citenamefont {Others}}]{Ludlow2008}%
  \BibitemOpen
  \bibfield  {author} {\bibinfo {author} {\bibfnamefont {A.~D.}\ \bibnamefont
  {Ludlow}}, \bibinfo {author} {\bibfnamefont {T.}~\bibnamefont {Zelevinsky}},
  \bibinfo {author} {\bibfnamefont {G.~K.}\ \bibnamefont {Campbell}}, \bibinfo
  {author} {\bibfnamefont {S.}~\bibnamefont {Blatt}}, \bibinfo {author}
  {\bibfnamefont {M.~M.}\ \bibnamefont {Boyd}}, \bibinfo {author}
  {\bibfnamefont {M.~H.~G.}\ \bibnamefont {{De Miranda}}}, \bibinfo {author}
  {\bibfnamefont {M.~J.}\ \bibnamefont {Martin}}, \bibinfo {author}
  {\bibfnamefont {J.~W.}\ \bibnamefont {Thomsen}}, \bibinfo {author}
  {\bibfnamefont {S.~M.}\ \bibnamefont {Foreman}}, \bibinfo {author}
  {\bibfnamefont {J.}~\bibnamefont {Ye}}, \ and\ \bibinfo {author}
  {\bibnamefont {Others}},\ }\href@noop {} {\bibfield  {journal} {\bibinfo
  {journal} {Science}\ }\textbf {\bibinfo {volume} {319}},\ \bibinfo {pages}
  {1805} (\bibinfo {year} {2008})}\BibitemShut {NoStop}%
\bibitem [{\citenamefont {Chou}\ \emph {et~al.}(2010)\citenamefont {Chou},
  \citenamefont {Hume}, \citenamefont {Koelemeij}, \citenamefont {Wineland},\
  and\ \citenamefont {Rosenband}}]{Chou2010}%
  \BibitemOpen
  \bibfield  {author} {\bibinfo {author} {\bibfnamefont {C.~W.}\ \bibnamefont
  {Chou}}, \bibinfo {author} {\bibfnamefont {D.~B.}\ \bibnamefont {Hume}},
  \bibinfo {author} {\bibfnamefont {J.~C.~J.}\ \bibnamefont {Koelemeij}},
  \bibinfo {author} {\bibfnamefont {D.~J.}\ \bibnamefont {Wineland}}, \ and\
  \bibinfo {author} {\bibfnamefont {T.}~\bibnamefont {Rosenband}},\ }\href
  {\doibase 10.1103/PhysRevLett.104.070802} {\bibfield  {journal} {\bibinfo
  {journal} {Phys. Rev. Lett.}\ }\textbf {\bibinfo {volume} {104}},\ \bibinfo
  {pages} {070802} (\bibinfo {year} {2010})}\BibitemShut {NoStop}%
\bibitem [{\citenamefont {Jiang}\ \emph {et~al.}(2011)\citenamefont {Jiang},
  \citenamefont {Ludlow}, \citenamefont {Lemke}, \citenamefont {Fox},
  \citenamefont {Sherman}, \citenamefont {Ma},\ and\ \citenamefont
  {Oates}}]{Jiang2011Yb_opt_lattice}%
  \BibitemOpen
  \bibfield  {author} {\bibinfo {author} {\bibfnamefont {Y.~Y.}\ \bibnamefont
  {Jiang}}, \bibinfo {author} {\bibfnamefont {A.~D.}\ \bibnamefont {Ludlow}},
  \bibinfo {author} {\bibfnamefont {N.~D.}\ \bibnamefont {Lemke}}, \bibinfo
  {author} {\bibfnamefont {R.~W.}\ \bibnamefont {Fox}}, \bibinfo {author}
  {\bibfnamefont {J.~A.}\ \bibnamefont {Sherman}}, \bibinfo {author}
  {\bibfnamefont {L.-S.}\ \bibnamefont {Ma}}, \ and\ \bibinfo {author}
  {\bibfnamefont {C.~W.}\ \bibnamefont {Oates}},\ }\href@noop {} {\bibfield
  {journal} {\bibinfo  {journal} {Nat. Photonics}\ }\textbf {\bibinfo {volume}
  {5}},\ \bibinfo {pages} {158} (\bibinfo {year} {2011})}\BibitemShut {NoStop}%
\bibitem [{\citenamefont {Derevianko}\ and\ \citenamefont
  {Katori}(2011)}]{Derevianko2011C}%
  \BibitemOpen
  \bibfield  {author} {\bibinfo {author} {\bibfnamefont {A.}~\bibnamefont
  {Derevianko}}\ and\ \bibinfo {author} {\bibfnamefont {H.}~\bibnamefont
  {Katori}},\ }\href@noop {} {\bibfield  {journal} {\bibinfo  {journal} {Rev.
  Mod. Phys.}\ }\textbf {\bibinfo {volume} {83}},\ \bibinfo {pages} {331}
  (\bibinfo {year} {2011})}\BibitemShut {NoStop}%
\bibitem [{\citenamefont {Misner}\ \emph {et~al.}(1973)\citenamefont {Misner},
  \citenamefont {Thorne},\ and\ \citenamefont {Wheeler}}]{Misner_Gravitation}%
  \BibitemOpen
  \bibfield  {author} {\bibinfo {author} {\bibfnamefont {C.~W.}\ \bibnamefont
  {Misner}}, \bibinfo {author} {\bibfnamefont {K.~S.}\ \bibnamefont {Thorne}},
  \ and\ \bibinfo {author} {\bibfnamefont {J.~A.}\ \bibnamefont {Wheeler}},\
  }\href@noop {} {\emph {\bibinfo {title} {{Gravitation}}}}\ (\bibinfo
  {publisher} {Freeman},\ \bibinfo {address} {New York},\ \bibinfo {year}
  {1973})\BibitemShut {NoStop}%
\bibitem [{\citenamefont {Zangwill}(2013)}]{Zangwill_MED}%
  \BibitemOpen
  \bibfield  {author} {\bibinfo {author} {\bibfnamefont {A.}~\bibnamefont
  {Zangwill}},\ }\href@noop {} {\emph {\bibinfo {title} {{Modern
  Electrodynamics}}}}\ (\bibinfo  {publisher} {Cambridge University Press},\
  \bibinfo {address} {Cambridge},\ \bibinfo {year} {2013})\BibitemShut
  {NoStop}%
\bibitem [{\citenamefont {Flambaum}\ and\ \citenamefont
  {Tedesco}(2006)}]{Flambaum2006A}%
  \BibitemOpen
  \bibfield  {author} {\bibinfo {author} {\bibfnamefont {V.~V.}\ \bibnamefont
  {Flambaum}}\ and\ \bibinfo {author} {\bibfnamefont {A.~F.}\ \bibnamefont
  {Tedesco}},\ }\href@noop {} {\bibfield  {journal} {\bibinfo  {journal} {Phys.
  Rev. C}\ }\textbf {\bibinfo {volume} {73}},\ \bibinfo {pages} {055501}
  (\bibinfo {year} {2006})}\BibitemShut {NoStop}%
\bibitem [{\citenamefont {Dinh}\ \emph {et~al.}(2009)\citenamefont {Dinh},
  \citenamefont {Dunning}, \citenamefont {Dzuba},\ and\ \citenamefont
  {Flambaum}}]{Dinh2009}%
  \BibitemOpen
  \bibfield  {author} {\bibinfo {author} {\bibfnamefont {T.~H.}\ \bibnamefont
  {Dinh}}, \bibinfo {author} {\bibfnamefont {A.}~\bibnamefont {Dunning}},
  \bibinfo {author} {\bibfnamefont {V.~A.}\ \bibnamefont {Dzuba}}, \ and\
  \bibinfo {author} {\bibfnamefont {V.~V.}\ \bibnamefont {Flambaum}},\
  }\href@noop {} {\bibfield  {journal} {\bibinfo  {journal} {Phys. Rev. A}\
  }\textbf {\bibinfo {volume} {79}},\ \bibinfo {pages} {054102} (\bibinfo
  {year} {2009})}\BibitemShut {NoStop}%
\bibitem [{\citenamefont {Berengut}\ and\ \citenamefont
  {Flambaum}(2011)}]{Berengut2011a}%
  \BibitemOpen
  \bibfield  {author} {\bibinfo {author} {\bibfnamefont {J.~C.}\ \bibnamefont
  {Berengut}}\ and\ \bibinfo {author} {\bibfnamefont {V.~V.}\ \bibnamefont
  {Flambaum}},\ }\href {\doibase 10.1088/1742-6596/264/1/012010} {\bibfield
  {journal} {\bibinfo  {journal} {J. Phys. Conf. Ser.}\ }\textbf {\bibinfo
  {volume} {264}} (\bibinfo {year} {2011}),\
  10.1088/1742-6596/264/1/012010}\BibitemShut {NoStop}%
\bibitem [{\citenamefont {Guena}\ \emph {et~al.}(2012)\citenamefont {Guena},
  \citenamefont {Abgrall}, \citenamefont {Rovera}, \citenamefont {Rosenbusch},
  \citenamefont {Tobar}, \citenamefont {Laurent}, \citenamefont {Clairon},\
  and\ \citenamefont {Bize}}]{Guena2012}%
  \BibitemOpen
  \bibfield  {author} {\bibinfo {author} {\bibfnamefont {J.}~\bibnamefont
  {Guena}}, \bibinfo {author} {\bibfnamefont {M.}~\bibnamefont {Abgrall}},
  \bibinfo {author} {\bibfnamefont {D.}~\bibnamefont {Rovera}}, \bibinfo
  {author} {\bibfnamefont {P.}~\bibnamefont {Rosenbusch}}, \bibinfo {author}
  {\bibfnamefont {M.~E.}\ \bibnamefont {Tobar}}, \bibinfo {author}
  {\bibfnamefont {P.}~\bibnamefont {Laurent}}, \bibinfo {author} {\bibfnamefont
  {A.}~\bibnamefont {Clairon}}, \ and\ \bibinfo {author} {\bibfnamefont
  {S.}~\bibnamefont {Bize}},\ }\href@noop {} {\bibfield  {journal} {\bibinfo
  {journal} {Phys. Rev. Lett.}\ }\textbf {\bibinfo {volume} {109}},\ \bibinfo
  {pages} {080801} (\bibinfo {year} {2012})}\BibitemShut {NoStop}%
\bibitem [{\citenamefont {Flambaum}(2006)}]{Flambaum2006C}%
  \BibitemOpen
  \bibfield  {author} {\bibinfo {author} {\bibfnamefont {V.~V.}\ \bibnamefont
  {Flambaum}},\ }\href@noop {} {\bibfield  {journal} {\bibinfo  {journal}
  {Phys. Rev. A}\ }\textbf {\bibinfo {volume} {73}},\ \bibinfo {pages} {034101}
  (\bibinfo {year} {2006})}\BibitemShut {NoStop}%
\bibitem [{\citenamefont {Flambaum}\ and\ \citenamefont
  {Kozlov}(2007)}]{Flambaum2007D}%
  \BibitemOpen
  \bibfield  {author} {\bibinfo {author} {\bibfnamefont {V.~V.}\ \bibnamefont
  {Flambaum}}\ and\ \bibinfo {author} {\bibfnamefont {M.~G.}\ \bibnamefont
  {Kozlov}},\ }\href {\doibase 10.1103/PhysRevLett.99.150801} {\bibfield
  {journal} {\bibinfo  {journal} {Phys. Rev. Lett.}\ }\textbf {\bibinfo
  {volume} {99}},\ \bibinfo {pages} {150801} (\bibinfo {year}
  {2007})}\BibitemShut {NoStop}%
\bibitem [{\citenamefont {DeMille}\ \emph {et~al.}(2008)\citenamefont
  {DeMille}, \citenamefont {Sainis}, \citenamefont {Sage}, \citenamefont
  {Bergeman}, \citenamefont {Kotochigova},\ and\ \citenamefont
  {Tiesinga}}]{DeMille2008A}%
  \BibitemOpen
  \bibfield  {author} {\bibinfo {author} {\bibfnamefont {D.}~\bibnamefont
  {DeMille}}, \bibinfo {author} {\bibfnamefont {S.}~\bibnamefont {Sainis}},
  \bibinfo {author} {\bibfnamefont {J.}~\bibnamefont {Sage}}, \bibinfo {author}
  {\bibfnamefont {T.}~\bibnamefont {Bergeman}}, \bibinfo {author}
  {\bibfnamefont {S.}~\bibnamefont {Kotochigova}}, \ and\ \bibinfo {author}
  {\bibfnamefont {E.}~\bibnamefont {Tiesinga}},\ }\href@noop {} {\bibfield
  {journal} {\bibinfo  {journal} {Phys. Rev. Lett.}\ }\textbf {\bibinfo
  {volume} {100}},\ \bibinfo {pages} {043202} (\bibinfo {year}
  {2008})}\BibitemShut {NoStop}%
\bibitem [{\citenamefont {Zelevinsky}\ \emph {et~al.}(2008)\citenamefont
  {Zelevinsky}, \citenamefont {Kotochigova},\ and\ \citenamefont
  {Ye}}]{Zelevinsky2008A}%
  \BibitemOpen
  \bibfield  {author} {\bibinfo {author} {\bibfnamefont {T.}~\bibnamefont
  {Zelevinsky}}, \bibinfo {author} {\bibfnamefont {S.}~\bibnamefont
  {Kotochigova}}, \ and\ \bibinfo {author} {\bibfnamefont {J.}~\bibnamefont
  {Ye}},\ }\href {\doibase 10.1103/PhysRevLett.100.043201} {\bibfield
  {journal} {\bibinfo  {journal} {Phys. Rev. Lett.}\ }\textbf {\bibinfo
  {volume} {100}},\ \bibinfo {pages} {043201} (\bibinfo {year}
  {2008})}\BibitemShut {NoStop}%
\bibitem [{\citenamefont {Kozlov}(2009)}]{Kozlov2009A}%
  \BibitemOpen
  \bibfield  {author} {\bibinfo {author} {\bibfnamefont {M.~G.}\ \bibnamefont
  {Kozlov}},\ }\href@noop {} {\bibfield  {journal} {\bibinfo  {journal} {Phys.
  Rev. A}\ }\textbf {\bibinfo {volume} {80}},\ \bibinfo {pages} {022118}
  (\bibinfo {year} {2009})}\BibitemShut {NoStop}%
\bibitem [{\citenamefont {Kozlov}(2013)}]{Kozlov2013B}%
  \BibitemOpen
  \bibfield  {author} {\bibinfo {author} {\bibfnamefont {M.~G.}\ \bibnamefont
  {Kozlov}},\ }\href@noop {} {\bibfield  {journal} {\bibinfo  {journal} {Phys.
  Rev. A}\ }\textbf {\bibinfo {volume} {87}},\ \bibinfo {pages} {032104}
  (\bibinfo {year} {2013})}\BibitemShut {NoStop}%
\bibitem [{\citenamefont {Kozlov}\ and\ \citenamefont
  {Levshakov}(2013)}]{Kozlov2013C}%
  \BibitemOpen
  \bibfield  {author} {\bibinfo {author} {\bibfnamefont {M.~G.}\ \bibnamefont
  {Kozlov}}\ and\ \bibinfo {author} {\bibfnamefont {S.~A.}\ \bibnamefont
  {Levshakov}},\ }\href {\doibase 10.1002/andp.201300010} {\bibfield  {journal}
  {\bibinfo  {journal} {Ann. Phys.}\ }\textbf {\bibinfo {volume} {525}},\
  \bibinfo {pages} {452} (\bibinfo {year} {2013})}\BibitemShut {NoStop}%
\bibitem [{\citenamefont {Flambaum}\ \emph {et~al.}(2013)\citenamefont
  {Flambaum}, \citenamefont {Stadnik}, \citenamefont {Kozlov},\ and\
  \citenamefont {Petrov}}]{Flambaum2013B}%
  \BibitemOpen
  \bibfield  {author} {\bibinfo {author} {\bibfnamefont {V.~V.}\ \bibnamefont
  {Flambaum}}, \bibinfo {author} {\bibfnamefont {Y.~V.}\ \bibnamefont
  {Stadnik}}, \bibinfo {author} {\bibfnamefont {M.~G.}\ \bibnamefont {Kozlov}},
  \ and\ \bibinfo {author} {\bibfnamefont {A.~N.}\ \bibnamefont {Petrov}},\
  }\href@noop {} {\bibfield  {journal} {\bibinfo  {journal} {Phys. Rev. A}\
  }\textbf {\bibinfo {volume} {88}},\ \bibinfo {pages} {052124} (\bibinfo
  {year} {2013})}\BibitemShut {NoStop}%
\bibitem [{\citenamefont {Jefferts}(2013)}]{Clocks_comp_2013}%
  \BibitemOpen
  \bibfield  {author} {\bibinfo {author} {\bibfnamefont {S.~R.}\ \bibnamefont
  {Jefferts}},\ }\href
  {http://www-conf.slac.stanford.edu/lsow/2013-presentations/SJefferts-AtomicClocks.pdf}
  {\emph {\bibinfo {title} {Atomic Clocks: Primary Frequency Standards at
  NIST}}}\ (\bibinfo {address} {8th Annual DOE Laser Safety Officer Workshop},\
  \bibinfo {year} {2013})\BibitemShut {NoStop}%
\bibitem [{\citenamefont {Stadnik}\ and\ \citenamefont
  {Flambaum}(2014)}]{Stadnik2014}%
  \BibitemOpen
  \bibfield  {author} {\bibinfo {author} {\bibfnamefont {Y.~V.}\ \bibnamefont
  {Stadnik}}\ and\ \bibinfo {author} {\bibfnamefont {V.~V.}\ \bibnamefont
  {Flambaum}},\ }\href {\doibase 10.1103/PhysRevD.89.043522} {\bibfield
  {journal} {\bibinfo  {journal} {Phys. Rev. D}\ }\textbf {\bibinfo {volume}
  {89}},\ \bibinfo {pages} {043522} (\bibinfo {year} {2014})}\BibitemShut
  {NoStop}%
\bibitem [{\citenamefont {Roberts}\ \emph {et~al.}(2014)\citenamefont
  {Roberts}, \citenamefont {Stadnik}, \citenamefont {Dzuba}, \citenamefont
  {Flambaum}, \citenamefont {Leefer},\ and\ \citenamefont
  {Budker}}]{Roberts2014}%
  \BibitemOpen
  \bibfield  {author} {\bibinfo {author} {\bibfnamefont {B.~M.}\ \bibnamefont
  {Roberts}}, \bibinfo {author} {\bibfnamefont {Y.~V.}\ \bibnamefont
  {Stadnik}}, \bibinfo {author} {\bibfnamefont {V.~A.}\ \bibnamefont {Dzuba}},
  \bibinfo {author} {\bibfnamefont {V.~V.}\ \bibnamefont {Flambaum}}, \bibinfo
  {author} {\bibfnamefont {N.}~\bibnamefont {Leefer}}, \ and\ \bibinfo {author}
  {\bibfnamefont {D.}~\bibnamefont {Budker}},\ }\href@noop {} {\bibfield
  {journal} {\bibinfo  {journal} {(Accepted to Phys. Rev. Lett.);
  arXiv:1404.2723.}\ } (\bibinfo {year} {2014})}\BibitemShut {NoStop}%
\bibitem [{\citenamefont {Baker}\ \emph {et~al.}(2006)\citenamefont {Baker},
  \citenamefont {Doyle}, \citenamefont {Geltenbort}, \citenamefont {Green},
  \citenamefont {van~der Grinten}, \citenamefont {Harris}, \citenamefont
  {Iaydjiev}, \citenamefont {Ivanov}, \citenamefont {May}, \citenamefont
  {Pendlebury}, \citenamefont {Richardson}, \citenamefont {Shiers},\ and\
  \citenamefont {Smith}}]{Baker2006NEDM}%
  \BibitemOpen
  \bibfield  {author} {\bibinfo {author} {\bibfnamefont {C.~A.}\ \bibnamefont
  {Baker}}, \bibinfo {author} {\bibfnamefont {D.~D.}\ \bibnamefont {Doyle}},
  \bibinfo {author} {\bibfnamefont {P.}~\bibnamefont {Geltenbort}}, \bibinfo
  {author} {\bibfnamefont {K.}~\bibnamefont {Green}}, \bibinfo {author}
  {\bibfnamefont {M.~G.~D.}\ \bibnamefont {van~der Grinten}}, \bibinfo {author}
  {\bibfnamefont {P.~G.}\ \bibnamefont {Harris}}, \bibinfo {author}
  {\bibfnamefont {P.}~\bibnamefont {Iaydjiev}}, \bibinfo {author}
  {\bibfnamefont {S.~N.}\ \bibnamefont {Ivanov}}, \bibinfo {author}
  {\bibfnamefont {D.~J.~R.}\ \bibnamefont {May}}, \bibinfo {author}
  {\bibfnamefont {J.~M.}\ \bibnamefont {Pendlebury}}, \bibinfo {author}
  {\bibfnamefont {J.~D.}\ \bibnamefont {Richardson}}, \bibinfo {author}
  {\bibfnamefont {D.}~\bibnamefont {Shiers}}, \ and\ \bibinfo {author}
  {\bibfnamefont {K.~F.}\ \bibnamefont {Smith}},\ }\href {\doibase
  10.1103/PhysRevLett.97.131801} {\bibfield  {journal} {\bibinfo  {journal}
  {Phys. Rev. Lett.}\ }\textbf {\bibinfo {volume} {97}},\ \bibinfo {pages}
  {131801} (\bibinfo {year} {2006})}\BibitemShut {NoStop}%
\bibitem [{\citenamefont {Griffith}\ \emph {et~al.}(2009)\citenamefont
  {Griffith}, \citenamefont {Swallows}, \citenamefont {Loftus}, \citenamefont
  {Romalis}, \citenamefont {Heckel},\ and\ \citenamefont
  {Fortson}}]{Griffith2009improved}%
  \BibitemOpen
  \bibfield  {author} {\bibinfo {author} {\bibfnamefont {W.~C.}\ \bibnamefont
  {Griffith}}, \bibinfo {author} {\bibfnamefont {M.~D.}\ \bibnamefont
  {Swallows}}, \bibinfo {author} {\bibfnamefont {T.~H.}\ \bibnamefont
  {Loftus}}, \bibinfo {author} {\bibfnamefont {M.~V.}\ \bibnamefont {Romalis}},
  \bibinfo {author} {\bibfnamefont {B.~R.}\ \bibnamefont {Heckel}}, \ and\
  \bibinfo {author} {\bibfnamefont {E.~N.}\ \bibnamefont {Fortson}},\
  }\href@noop {} {\bibfield  {journal} {\bibinfo  {journal} {Phys. Rev. Lett.}\
  }\textbf {\bibinfo {volume} {102}},\ \bibinfo {pages} {101601} (\bibinfo
  {year} {2009})}\BibitemShut {NoStop}%
\bibitem [{\citenamefont {Swallows}\ \emph {et~al.}(2013)\citenamefont
  {Swallows}, \citenamefont {Loftus}, \citenamefont {Griffith}, \citenamefont
  {Heckel}, \citenamefont {Fortson},\ and\ \citenamefont
  {Romalis}}]{Swallows2013}%
  \BibitemOpen
  \bibfield  {author} {\bibinfo {author} {\bibfnamefont {M.~D.}\ \bibnamefont
  {Swallows}}, \bibinfo {author} {\bibfnamefont {T.~H.}\ \bibnamefont
  {Loftus}}, \bibinfo {author} {\bibfnamefont {W.~C.}\ \bibnamefont
  {Griffith}}, \bibinfo {author} {\bibfnamefont {B.~R.}\ \bibnamefont
  {Heckel}}, \bibinfo {author} {\bibfnamefont {E.~N.}\ \bibnamefont {Fortson}},
  \ and\ \bibinfo {author} {\bibfnamefont {M.~V.}\ \bibnamefont {Romalis}},\
  }\href {\doibase 10.1103/PhysRevA.87.012102} {\bibfield  {journal} {\bibinfo
  {journal} {Phys. Rev. A}\ }\textbf {\bibinfo {volume} {87}},\ \bibinfo
  {pages} {012102} (\bibinfo {year} {2013})}\BibitemShut {NoStop}%
\bibitem [{\citenamefont {Rosenberry}\ and\ \citenamefont
  {Chupp}(2001)}]{Rosenberry2001atomic}%
  \BibitemOpen
  \bibfield  {author} {\bibinfo {author} {\bibfnamefont {M.~A.}\ \bibnamefont
  {Rosenberry}}\ and\ \bibinfo {author} {\bibfnamefont {T.~E.}\ \bibnamefont
  {Chupp}},\ }\href@noop {} {\bibfield  {journal} {\bibinfo  {journal} {Phys.
  Rev. Lett.}\ }\textbf {\bibinfo {volume} {86}},\ \bibinfo {pages} {22}
  (\bibinfo {year} {2001})}\BibitemShut {NoStop}%
\bibitem [{\citenamefont {Player}\ and\ \citenamefont
  {Sandars}(1970)}]{Player1970experiment}%
  \BibitemOpen
  \bibfield  {author} {\bibinfo {author} {\bibfnamefont {M.~A.}\ \bibnamefont
  {Player}}\ and\ \bibinfo {author} {\bibfnamefont {P.~G.~H.}\ \bibnamefont
  {Sandars}},\ }\href@noop {} {\bibfield  {journal} {\bibinfo  {journal} {J.
  Phys. B At. Mol. Phys.}\ }\textbf {\bibinfo {volume} {3}},\ \bibinfo {pages}
  {1620} (\bibinfo {year} {1970})}\BibitemShut {NoStop}%
\bibitem [{\citenamefont {Ensberg}(1967)}]{Ensberg1967}%
  \BibitemOpen
  \bibfield  {author} {\bibinfo {author} {\bibfnamefont {E.~S.}\ \bibnamefont
  {Ensberg}},\ }\href@noop {} {\bibfield  {journal} {\bibinfo  {journal} {Phys.
  Rev.}\ }\textbf {\bibinfo {volume} {153}},\ \bibinfo {pages} {36} (\bibinfo
  {year} {1967})}\BibitemShut {NoStop}%
\bibitem [{\citenamefont {Murthy}\ \emph {et~al.}(1989)\citenamefont {Murthy},
  \citenamefont {{Krause Jr}}, \citenamefont {Li},\ and\ \citenamefont
  {Hunter}}]{Murthy1989new}%
  \BibitemOpen
  \bibfield  {author} {\bibinfo {author} {\bibfnamefont {S.~A.}\ \bibnamefont
  {Murthy}}, \bibinfo {author} {\bibfnamefont {D.}~\bibnamefont {{Krause Jr}}},
  \bibinfo {author} {\bibfnamefont {Z.~L.}\ \bibnamefont {Li}}, \ and\ \bibinfo
  {author} {\bibfnamefont {L.~R.}\ \bibnamefont {Hunter}},\ }\href@noop {}
  {\bibfield  {journal} {\bibinfo  {journal} {Phys. Rev. Lett.}\ }\textbf
  {\bibinfo {volume} {63}},\ \bibinfo {pages} {965} (\bibinfo {year}
  {1989})}\BibitemShut {NoStop}%
\bibitem [{\citenamefont {Regan}\ \emph {et~al.}(2002)\citenamefont {Regan},
  \citenamefont {Commins}, \citenamefont {Schmidt},\ and\ \citenamefont
  {DeMille}}]{Regan2002new}%
  \BibitemOpen
  \bibfield  {author} {\bibinfo {author} {\bibfnamefont {B.~C.}\ \bibnamefont
  {Regan}}, \bibinfo {author} {\bibfnamefont {E.~D.}\ \bibnamefont {Commins}},
  \bibinfo {author} {\bibfnamefont {C.~J.}\ \bibnamefont {Schmidt}}, \ and\
  \bibinfo {author} {\bibfnamefont {D.}~\bibnamefont {DeMille}},\ }\href@noop
  {} {\bibfield  {journal} {\bibinfo  {journal} {Phys. Rev. Lett.}\ }\textbf
  {\bibinfo {volume} {88}},\ \bibinfo {pages} {071805} (\bibinfo {year}
  {2002})}\BibitemShut {NoStop}%
\bibitem [{\citenamefont {Hudson}\ \emph {et~al.}(2011)\citenamefont {Hudson},
  \citenamefont {Kara}, \citenamefont {Smallman}, \citenamefont {Sauer},
  \citenamefont {Tarbutt},\ and\ \citenamefont {Hinds}}]{Hudson2011improved}%
  \BibitemOpen
  \bibfield  {author} {\bibinfo {author} {\bibfnamefont {J.~J.}\ \bibnamefont
  {Hudson}}, \bibinfo {author} {\bibfnamefont {D.~M.}\ \bibnamefont {Kara}},
  \bibinfo {author} {\bibfnamefont {I.~J.}\ \bibnamefont {Smallman}}, \bibinfo
  {author} {\bibfnamefont {B.~E.}\ \bibnamefont {Sauer}}, \bibinfo {author}
  {\bibfnamefont {M.~R.}\ \bibnamefont {Tarbutt}}, \ and\ \bibinfo {author}
  {\bibfnamefont {E.~A.}\ \bibnamefont {Hinds}},\ }\href@noop {} {\bibfield
  {journal} {\bibinfo  {journal} {Nature}\ }\textbf {\bibinfo {volume} {473}},\
  \bibinfo {pages} {493} (\bibinfo {year} {2011})}\BibitemShut {NoStop}%
\bibitem [{\citenamefont {Kara}\ \emph {et~al.}(2012)\citenamefont {Kara},
  \citenamefont {Smallman}, \citenamefont {Hudson}, \citenamefont {Sauer},
  \citenamefont {Tarbutt},\ and\ \citenamefont {Hinds}}]{Kara2012measurement}%
  \BibitemOpen
  \bibfield  {author} {\bibinfo {author} {\bibfnamefont {D.~M.}\ \bibnamefont
  {Kara}}, \bibinfo {author} {\bibfnamefont {I.~J.}\ \bibnamefont {Smallman}},
  \bibinfo {author} {\bibfnamefont {J.~J.}\ \bibnamefont {Hudson}}, \bibinfo
  {author} {\bibfnamefont {B.~E.}\ \bibnamefont {Sauer}}, \bibinfo {author}
  {\bibfnamefont {M.~R.}\ \bibnamefont {Tarbutt}}, \ and\ \bibinfo {author}
  {\bibfnamefont {E.~A.}\ \bibnamefont {Hinds}},\ }\href@noop {} {\bibfield
  {journal} {\bibinfo  {journal} {New J. Phys.}\ }\textbf {\bibinfo {volume}
  {14}},\ \bibinfo {pages} {103051} (\bibinfo {year} {2012})}\BibitemShut
  {NoStop}%
\bibitem [{\citenamefont {Baron}\ \emph {et~al.}(2014)\citenamefont {Baron},
  \citenamefont {Campbell}, \citenamefont {DeMille}, \citenamefont {Doyle},
  \citenamefont {Gabrielse}, \citenamefont {Gurevich}, \citenamefont {Hess},
  \citenamefont {Hutzler}, \citenamefont {Kirilov}, \citenamefont {Kozyryev},
  \citenamefont {O'Leary}, \citenamefont {Panda}, \citenamefont {Parsons},
  \citenamefont {Petrik}, \citenamefont {Spaun}, \citenamefont {Vutha},\ and\
  \citenamefont {West}}]{Baron2014}%
  \BibitemOpen
  \bibfield  {author} {\bibinfo {author} {\bibfnamefont {J.}~\bibnamefont
  {Baron}}, \bibinfo {author} {\bibfnamefont {W.~C.}\ \bibnamefont {Campbell}},
  \bibinfo {author} {\bibfnamefont {D.}~\bibnamefont {DeMille}}, \bibinfo
  {author} {\bibfnamefont {J.~M.}\ \bibnamefont {Doyle}}, \bibinfo {author}
  {\bibfnamefont {G.}~\bibnamefont {Gabrielse}}, \bibinfo {author}
  {\bibfnamefont {Y.~V.}\ \bibnamefont {Gurevich}}, \bibinfo {author}
  {\bibfnamefont {P.~W.}\ \bibnamefont {Hess}}, \bibinfo {author}
  {\bibfnamefont {N.~R.}\ \bibnamefont {Hutzler}}, \bibinfo {author}
  {\bibfnamefont {E.}~\bibnamefont {Kirilov}}, \bibinfo {author} {\bibfnamefont
  {I.}~\bibnamefont {Kozyryev}}, \bibinfo {author} {\bibfnamefont {B.~R.}\
  \bibnamefont {O'Leary}}, \bibinfo {author} {\bibfnamefont {C.~D.}\
  \bibnamefont {Panda}}, \bibinfo {author} {\bibfnamefont {M.~F.}\ \bibnamefont
  {Parsons}}, \bibinfo {author} {\bibfnamefont {E.~S.}\ \bibnamefont {Petrik}},
  \bibinfo {author} {\bibfnamefont {B.}~\bibnamefont {Spaun}}, \bibinfo
  {author} {\bibfnamefont {A.~C.}\ \bibnamefont {Vutha}}, \ and\ \bibinfo
  {author} {\bibfnamefont {A.~D.}\ \bibnamefont {West}},\ }\href {\doibase
  10.1126/science.1248213} {\bibfield  {journal} {\bibinfo  {journal}
  {Science}\ }\textbf {\bibinfo {volume} {343}},\ \bibinfo {pages} {269}
  (\bibinfo {year} {2014})}\BibitemShut {NoStop}%
\bibitem [{\citenamefont {Eckel}\ \emph {et~al.}(2013)\citenamefont {Eckel},
  \citenamefont {Hamilton}, \citenamefont {Kirilov}, \citenamefont {Smith},\
  and\ \citenamefont {DeMille}}]{Eckel2013search}%
  \BibitemOpen
  \bibfield  {author} {\bibinfo {author} {\bibfnamefont {S.}~\bibnamefont
  {Eckel}}, \bibinfo {author} {\bibfnamefont {P.}~\bibnamefont {Hamilton}},
  \bibinfo {author} {\bibfnamefont {E.}~\bibnamefont {Kirilov}}, \bibinfo
  {author} {\bibfnamefont {H.~W.}\ \bibnamefont {Smith}}, \ and\ \bibinfo
  {author} {\bibfnamefont {D.}~\bibnamefont {DeMille}},\ }\href@noop {}
  {\bibfield  {journal} {\bibinfo  {journal} {Phys. Rev. A}\ }\textbf {\bibinfo
  {volume} {87}},\ \bibinfo {pages} {052130} (\bibinfo {year}
  {2013})}\BibitemShut {NoStop}%
\end{thebibliography}%

\end{document}